\documentclass{aa}  

\usepackage{graphicx}
\usepackage{txfonts}
\usepackage{threeparttable}
\usepackage{color}
\usepackage{array}
\usepackage{rotating}
\newcolumntype{H}{>{\setbox0=\hbox\bgroup}c<{\egroup}@{}}
\begin{document}

   \title{Extragalactic globular cluster near-infrared spectroscopy}

   \subtitle{I. Integrated near-infrared spectra of Centaurus\,A/NGC\,5128}

   \author{L. G. Dahmer-Hahn \inst{1} \and
           A. L. Chies-Santos\inst{2} \and
           E. Eftekhari\inst{3,4} \and
           E. Zanatta\inst{5} \and
           R. Riffel\inst{2,3} \and
           A. Vazdekis\inst{3,4} \and
           A. Villaume \inst{6,7} \and
           M. A. Beasley\inst{3,4,8} \and
           A. E. Lassen\inst{2,9}}

   \institute{Shanghai Astronomical Observatory, Chinese Academy of Sciences, 80 Nandan Road, Shanghai 200030, China\\ \email{luisgdh@shao.ac.cn} \and
              Departamento de Astronomia, Instituto de Física, Universidade Federal do Rio Grande do Sul (UFRGS), Av. Bento Gonçalves, 9500, Porto Alegre, RS, Brazil\\ \email{ana.chies@ufrgs.br} \and
              Instituto de Astrof\'isica de Canarias, E-38200 La Laguna, Tenerife, Spain \and
              Departamento de Astrof\'isica, Universidad de La Laguna, E-38205 La Laguna, Tenerife, Spain \and
              Instituto de Astronomia, Geof\'isica e Ci\^encias Atmosf\'ericas, Universidade de S\~ao Paulo, 05508-900 S\~ao Paulo, Brazil \and
              Waterloo Centre for Astrophysics, University of Waterloo, Waterloo, Ontario, N2L 3G1, Canada \and
              Department of Physics and Astronomy, University of Waterloo, Waterloo, Ontario N2L 3G1, Canada \and 
              Centre for Astrophysics and Supercomputing, Swinburne University, John Street, Hawthorn VIC 3122, Australia \and
              Department of Astronomy, The Oskar Klein Centre, Stockholm University, AlbaNova, SE-10691 Stockholm, Sweden}

   \date{Received September 15, 1996; accepted March 16, 1997}

% \abstract{}{}{}{}{} 
% 5 {} token are mandatory
 
  \abstract
   {One way to constrain the evolutionary histories of galaxies is to analyse their stellar populations. In the local Universe, our understanding of the stellar population properties of galaxies has traditionally relied on the study of optical absorption and emission-line features.}
   {In order to overcome limitations intrinsic to this wavelength range, such as the age-metallicity degeneracy and the high sensitivity to dust reddening, we must use wavelength ranges beyond the optical. The near-infrared (NIR) offers a possibility to extract information on spectral signatures that are not as obvious in traditional optical bands. Moreover, with the current and forthcoming generation of instrumentation focusing on the NIR, it is mandatory to explore possibilities within this wavelength range for nearby-Universe galaxies. However, although the NIR  shows great potential, we are only beginning to understand it. Widely used techniques such as a full spectral fitting and line strength indices need to be tested on systems that are as close to simple stellar populations as possible, and the result from the techniques need to be compared to the yields from a traditional optical analysis.}
   {We present a NIR spectral survey of extragalactic globular clusters (GCs). The set was composed of 21 GCs from the Centaurus\,A galaxy that were obtained with SOAR/TripleSpec4, which covered the $\sim$1.0-2.4\,$\mu$m range with a spectral resolution ($\rm{R={\lambda}/\Delta \lambda}$) of 3500. These spectra cover H$\beta$ equivalent widths between 0.98\,\r{A} and 4.32\,\r{A}, and [MgFe]' between 0.24\,\r{A} and 3.76\,\r{A}.}
   {This set was ideal for performing absorption band measurements and a full spectral fitting, and it can be used for kinematic studies and age and abundance measurements. With this library, we expect to be able to probe the capabilities of NIR models, as well as to further improve stellar population estimates for the GCs around the Centaurus~A galaxy.}
   {}


   \maketitle
%
%________________________________________________________________

\section{Introduction}
\label{sec:intro}
In order to understand how a galaxy has evolved, it is essential to determine its stellar population (SP) properties. With the star formation history (SFH) in hand, many different events during its evolution can then be traced, such as mergers \citep{Gorkom+86,Sesto+21}, past and/or ongoing interactions \citep{Abadi+99,Steyrleithner+20}, the presence or absence of bars and rings \citep{Allard+06,Sanchez-Blazquez+11,Azevedo+23}, an active galactic nucleus (AGN) and outflow activity \citep{Maiolino+17,Riffel+21,Riffel+22,luisgdh+22}, as well as gas and dust properties \citep{Dwek98,Riffel+09,Riffel+22,Li+21,Lassen+21,Reddy+22}.
\par
Although it is possible to individually resolve stars of very close sources \citep{Baade44,Brown+04, Brown+09,Crnojevic+16}, most galaxies beyond the Local Group need to be studied via integrated light. This analysis is either performed with photometry (with either broad- or narrow-band filters) or with spectroscopy (mainly performed through spectral index measurement or full-spectral fitting). With this purpose, the most commonly used wavelength range by far is the optical \citep[e.g.][among many others]{Worthey94, CF+04, CF+05, Beasley+08, Chen+10, Mallmann+18, Sanchez+16, luisgdh+22, Nascimento+22,Riffel+23}. However, despite their ubiquity, optical wavelengths pose many problems, such as the age-metallicity degeneracy \citep[which is stronger for colours, but still present for spectral index analysis and full-spectral fitting][]{Worthey94, Conroy+18}, the high sensitivity to dust reddening \citep{Riffel+08,Riffel+09}, and the low spatial resolution for ground-based telescopes.
\par
The near-infrared (NIR) is usually dominated by cool giants, which has long been proposed to soften or even resolve these problems when employed in combination with optical bands \citep{Maraston05, Marigo+08}. At the youngest ages, this domination of cool giants in the NIR is exacted through red supergiants, while at older ages, the red giant branch (RGB), red clump (RC), and asymptotic giant branch (AGB) dominate. For objects older than some billion years, the RGB/RC/AGB region is only weakly dependent on age. This means that it is possible to measure metallicities and abundances in a largely age-independent way. These abundances can be measured from a wealth of molecular bands in the NIR spectral region, including C$_2$, CN, CO, H$_2$, TiO, VO, FeH, and ZrO, as well as strong atomic lines for elements such as Na, Mg, Al, Si, K, Ca, Ti, and Fe \citep{Rayner+09,Riffel+07,Riffel+15,Riffel+19,Eftekhari+21}. The CO (in the NIR) and OH (at 3.3~$\mu$m) molecular features are of particular interest because they can be used to measure the O abundance, which is challenging to do in the optical with integrated light. Therefore, there is growing interest in studying stellar populations of nearby galaxies using the NIR wavelength range \citep{Origlia+93,Origlia&Oliva00,Martins+10,StorchiBergmann+12,Riffel+08,Riffel+09,RiffelRA+10,Riffel+11b,Riffel+15,Dametto+14,Dametto+19,Mason+15,Diniz+17,Diniz+19, luisgdh+18,luisgdh+19a, BaldwinC+18,Eftekhari+21,Riffel+19,Riffel+22,Riffel+24}.
\par
Although the first NIR spectra were obtained over 140 years ago \citep{Abney&Festing1881}, the instruments achieved a capacity comparable to optical spectroscopy only very recently. However, even though the NIR has the potential to resolve many issues in traditional optical studies, stellar population modelling in this wavelength range is still in its infancy, with significant differences in the predicted spectral energy distributions between models \citep{BC03, Maraston05, Conroy+09, Meneses-Goytia+15,Vazdekis+16,Millan-Irigoyen+21,Eftekhari+22a}. For example, \citet{BaldwinC+18} and \citet{luisgdh+18} analysed different sets of models and reported that current NIR models could not produce reliable results, with each library finding a different dominant population independent of the object spectra.
The most puzzling case is that of the CO absorption bands of massive early-type galaxies (ETGs), which are prominent in the H- and K-band spectral regions. This mismatch has so far been attributed to an enhanced contribution of stars in the thermally pulsating AGB evolutionary phase, which dominates the NIR light of intermediate-age \citep[peaking at $\sim$1.6~Gyr][]{Marigo+17} SPs. Whether this enhanced contribution is real is still a matter of debate \citep[see][for example]{Maraston05,Capozzi+16,Marigo+17,Eftekhari+22b}.
\par
\citet{Riffel+19} studied different libraries of models and found that the models fail to reproduce the observed stellar absorptions of galaxies. A few other important results \citep{Riffel+11a,Chies-Santos+11a,Chies-Santos+11b,Conroy+09,Powalka+16a,Powalka+16b,Dametto+19} also highlighted the issues of current simple stellar population (SSP) models to predict the properties of galactic and extragalactic globular clusters (GCs). Behind these issues is our lack of understanding of crucial phases that dominate the light in this range, such as thermally pulsing AGB stars.
\par
Empirical studies attempting to resolve these issues commonly make use of galaxy spectra, which are composed of many sets of stellar populations \citep[e.g.][]{Riffel+07, BaldwinC+18,luisgdh+18, Riffel+15,Riffel+19}. Ideally, these studies should be carried out with spectra of simpler stellar populations, such as GCs, in order to overcome these difficulties. However, the low number of public spectra for GCs outside the Milky Way is still an issue, especially in the NIR. Homogeneous and large samples of integrated spectra of GCs in the NIR would provide the opportunity to probe stellar populations with metallicities and ages beyond what is commonly found in the environment of our Galaxy. %One attempt at observing GCs in the NIR was made by \citet{Riffel+11a}, who obtained NIR spectroscopy of 12 GCs in our Galaxy. These spectra were analysed with \citet{Maraston05} models, and the authors found that these models can reproduce most of the observed absorptions, with the exception of \ion{Mg}{I}~$\lambda$1.49$\mu$m. They also found that, although models can reproduce the observed features in the optical, in the NIR they do underestimate the age of its stellar population.
\par
Although there are still many issues to be solved, the addition of the NIR spectral range to studies of GCs was shown to have great potential. For instance, \citet{Blakeslee+12,Chies-Santos+12,Cho+16} showed that by combining optical/NIR colour indices, the metallicities can be studied better than based on optical colours alone. 
At the same time, the NIR is nearly insensitive to the effect of hot horizontal branch stars, which are known to be present in massive old GCs \citep{Chies-Santos+12,Georgiev+12}.  
Nevertheless, although the stellar populations of GCs are simpler by far than galaxy populations, inconsistencies in stellar population measurements when using the NIR spectral range have persisted for about two decades.  \citet{Hempel+03} reported by directly comparing optical/NIR models to data large fractions of intermediate-age GCs ($\sim$ 2-8\,Gyr) in old elliptical galaxies. \citet{Chies-Santos+11a,Chies-Santos+11b} reported based on the analysis of a set of 14 galaxies using optical+NIR photometry no significant difference between the mean ages of GCs among elliptical galaxies.
%, but with S0 galaxies showing evidence for younger GCs.
%However, they 
They also reported that SSP models fail to reproduce GC properties in colour-colour diagrams. The offset between observed and modelled colours was likely to yield erroneous age results from colour-colour diagnostic diagrams. This result was later revisited and confirmed by \citet{Powalka+16b}, who analysed 2000 GCs from the Next Generation Virgo Survey \citep[NGVS][]{Ferrarese+12} using optical+NIR data. They reported that SSP models from different groups inconsistently match different colour-colour diagrams.
\citet{Powalka+16a} expanded these results and showed that the most popular SSP models used by the astronomical community disagree remarkably with the colour distributions of M87 GCs. The authors found that these variances can be attributed to environmental effects. A denser environment produces a wider dynamic range in certain colour indices. This demonstrates that stellar population diagnostics derived from model predictions calibrated on one particular sample of GCs (which is the case for most studies) may not be appropriate for all extragalactic GCs.
%in GC studies. For example, \citet{Georgiev+12} analyzed GCs of candidate intermediate-age early-type galaxies NGC 3610, 584 and 3377. They reported that a combination of optical+NIR is capable of disentangling the degeneracy between ages and metallicities between colours and SSP models, mainly since I-K' colour primarily measures a population's metallicity. At the same time, it is nearly insensitive to the effect of hot horizontal branch stars, known to be present in massive old GCs. 
\par 
Another important result was obtained by \citet{Usher+15}, who employed data from the SAGES Legacy Unifying Globulars and GalaxieS \citep[SLUGGS][]{Brodie+14}.
By using stacked GC spectra in the red part of the optical spectral range (6500 to 9000 \AA ), they showed that GCs with the same colours and absolute magnitudes in different galaxies showed different metal line strengths, suggesting different metallicities. 
After showing that the strength of the calcium triplet (CaT) spectral feature at 8600 \r{A} is largely insensitive to age for populations older than a few billion years \citep{Usher+19a}, \citet{Usher+19b} used a combination of optical photometry and metallicity measured from the CaT to show that the age distribution and age-metallicity relation of GCs can vary wildly between galaxies in a manner connected to the galactic assembly histories.
Thus, the stellar population properties of GCs seem to vary significantly for different galaxies \citep{Usher+15} and environments \citep{Powalka+16b}. 

\par
With the current \textit{James Webb} Space Telescope and upcoming telescopes and instruments focused on the NIR, such as the \textit{Euclid} and the \textit{Nancy Grace Roman} telescopes, and the planned NIR instruments on 30-40m class telescopes such as ELT\footnote{Extremely Large Telescope}/MOSAIC\footnote{Multi-Object Spectrograph for Astrophysics, Intergalactic-medium studies and Cosmology} and the next generation of spectrographs on the 10m class telescopes, that is, MOONS\footnote{Multi-Object Optical and Near-infrared Spectrograph}/VLT\footnote{Very Large Telescope} and PFS\footnote{Prime Focus Spectrograph}/Subaru, the need for reliable NIR SSP models especially for spectral analysis is clear. However, there is currently a lack of high-quality star cluster spectra in the NIR to verify stellar population models in this wavelength region. Previous GC libraries \citep[e.g.][]{Lyubenova+10,Sakari+16} have focused on a single NIR band at a time, which limits the amount of useful information and the number of comparisons that can be made.
\par
%GLOBALISE - GLOBulAr cLusters Infrared SpEctra
%EGlobIRS  - Extragalactic Globular Clusters with Infrared Spectroscopy
%LIBERTY   - extragaLactIc gloBular clustErs infRared specTroscopY
We present a spectral library of extragalactic globular clusters covering the whole 1.0-2.4$\mu$m at the same time. This is part of the extragalactic globular clusters near-infrared spectroscopy (LIBERTY) initiative, which is aimed at studying NIR spectra of GCs in nearby galaxies. Our interests are twofold. On the one hand, we wish to use GCs to test stellar population models, and on the other hand, we wish to study galaxy evolution with GCs. This study is the first of a series of papers analysing the properties of these GC spectra. The second paper of this series, \citet[][Paper~II]{Eftekhari+25}, describes the zero-point problem. Upcoming papers will focus on (i) CO indices, (ii) the remaining indices, and (iii) a full spectral fitting.
%(Paper III), and full spectral fitting (Paper IV).
\par
This paper is structured as follows: In Section~\ref{sec:data}, we describe our sample and observations, and we assess the data quality of our data. In Section~\ref{sec:results}, we present our preliminary results, and our final remarks are given in Section~\ref{sec:remarks}.

\section{Data}
\label{sec:data}
%Here we introduce the sample, the observations and data reduction procedures and assess the data quality.

\subsection{Sample selection}
\begin{figure*}
    \centering
    \includegraphics[width=\textwidth]{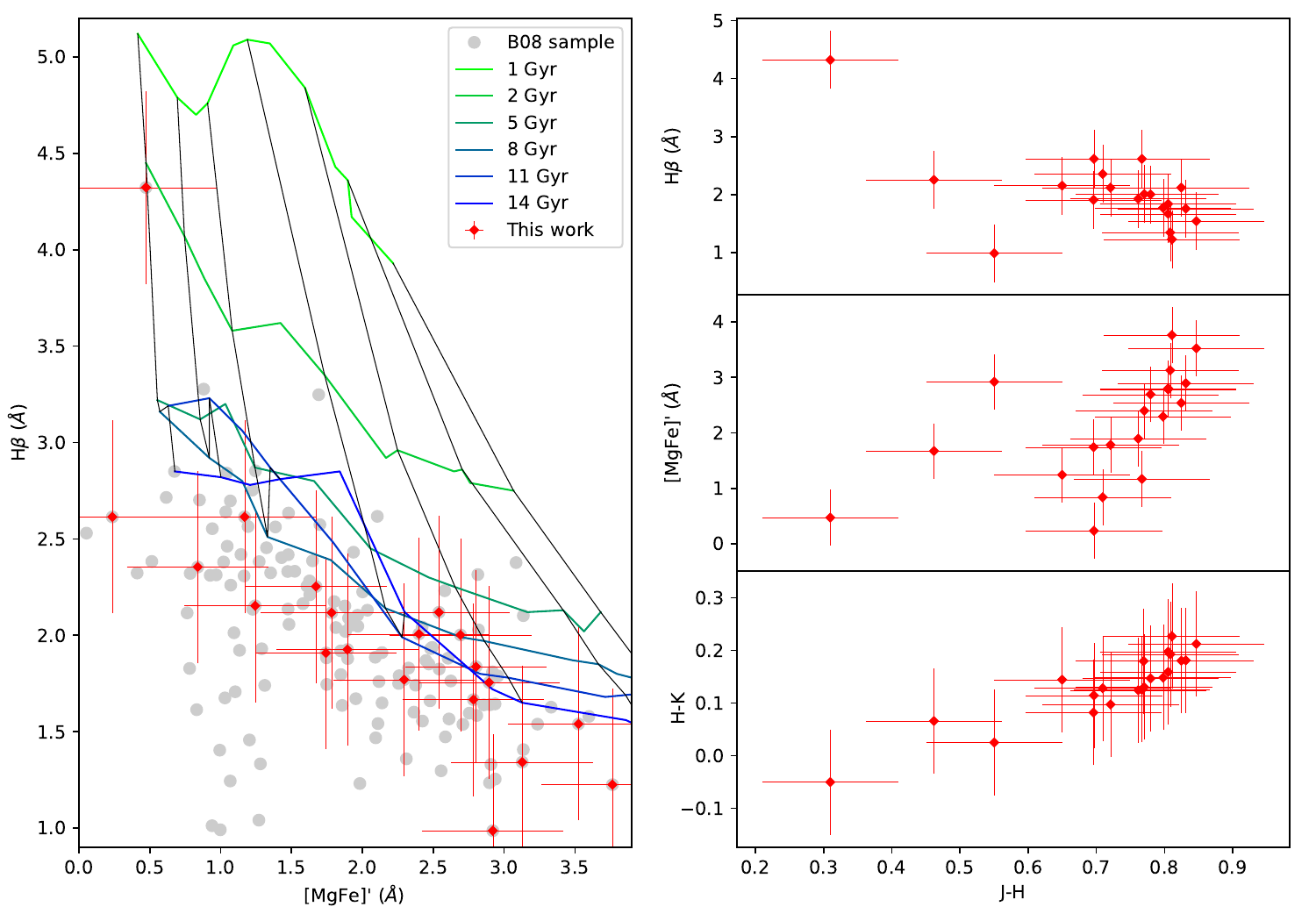}
    \caption{Main properties for our sample. In the {\it left} panel, we show that our sample (red diamonds) covers H$\beta$ and [MgFe]' values from B08 (grey circles). For comparison, we overplot the H$\beta\times$[MgFe]' grid measured from the E-MILES model library. Different ages are represented using different colours, and the different metallicities are all represented in black (from left to right, the [M/H] tracks are equal to -2.27, -1.79, -1.26, -0.66, -0.25,  +0.15, and +0.40). %, growing from subsolar (left) to supersolar (right). 
    In the {\it right} panels, we show the correlations between J-H colours (x-axis) and H$\beta$ ({\it upper}),  [MgFe]' ({\it middle}), and H-K colours ({\it bottom panels}). We plot the respective errors in all panels as red crosses over each respective diamond.}
    \label{fig:sample}
\end{figure*}

Our sample comprises 21 GCs of Centaurus~A/NGC\,5128 (hereafter Cen~A), which is one of the most massive galaxies in the nearby Universe. With a stellar mass of $\sim$10$^{11.21}M_\odot$, it lies at a distance of $\sim3.8$\,Mpc \citep{Harris+10,Fall+18}. These objects are a sub-sample of the objects presented by \citet[][hereafter B08]{Beasley+08}, which have publicly available optical spectra. In order to ensure that our sample covers as many ages and metallicities as possible, we used H$\beta$ as a proxy for age and [MgFe]' \citep[a composite index constructed from the combination of Mgb, Fe5270, and Fe5335 indices, see][]{Thomas+03} as a proxy for metallicity. We show in the left panel of Fig.~\ref{fig:sample} the H$\beta \times$[MgFe]' coverage in the original B08 sample. Our subsample is highlighted in red. 
\par
In order to emphasise the age and composition of our GCs, we overplot in the left panel of Fig~\ref{fig:sample} a grid with the measurements from the E-MILES stellar population library \citep{Rock+16,Vazdekis+16}. This grid was constructed using a \citet{Kroupa01} initial mass function and \citet{Pietrinferni+04} isochrones, spanning six ages (1, 2, 5, 8, 11, and 14~Gyr) and eight metallicities ([M/H]=-2.27, -1.79, -1.26, -0.66, -0.25,  +0.15, and +0.40). We discuss in Paper~II why most of B08 GCs appear to be older than the age of the Universe in this panel and how the NIR hydrogen index behaves in this regard.

% Out of those objects, we chose 21 GCs with optical S/N$>$30. To select our sample, we measured their Lick/IDS spectral indices \citep{Worthey+94}. We determined their ages and metallicities by fitting \citet{Lee&Worthey05} models with Monte-Carlo interactions. These results are plotted in Fig.\ref{fig:sample}, from where we selected our objects aimed at covering the widest possible age and metallicity range. Objects with S/N$>$30 are plotted in black, and our sub-sample of 21 objects is highlighted in red. Objects plotted in grey do not match our S/N threshold but are plotted for completeness.

\subsection{Observation and data reduction}

We obtained integrated spectra of these GCs with the TripleSpec4 instrument, which is attached to the Southern Astrophysical Research Telescope (SOAR) telescope, during two observing runs on 23 March and 24 April 2020, and on 12, 17, and 18 April 2021. The data were obtained through proposals SO2021A-016 (PI: L. G. Dahmer-Hahn) and SO2022A-009 (PI: E. Zanatta). The final sample is presented in Table~\ref{tab:sample}, together with the main properties.
\par
Since TripleSpec4 does not have any moving parts, all observations were performed with the default slit (1\farcs1), resulting in a spectral resolution of R\,$\simeq$\,3500 and 0\farcs33763/pixel sampling. The atmospheric seeing was not recorded for each individual object, but it varied between 0\farcs7 and 1\farcs2 throughout the observed nights. During daytime calibrations, we also took on and off flats for flat-field correction, and hollow lamps for the wavelength correction. However, we finally performed the wavelength calibration with sky lines because they were taken at the same time as the objects and therefore did not introduce time biases.
\par
We conducted the data reduction through the default {\sc Spextool idl} pipeline \citep{Vacca+03, Cushing+04}. This pipeline follows the standard NIR reduction process, which consists of flat-field correction, wavelength calibration (in a vacuum, calibrated from sky emission lines), spatial extraction, telluric band correction, and flux calibration. We performed the spectral extraction by integrating a region of 2\farcs0 centred on the flux peak ($\sim 3 \sigma$ extraction). 
\par
Due to the point-source nature of our sample, the observations were performed by nodding the objects along the slit (ABBA), which partially removed the sky emission. The remaining sky emission was removed by extracting a pure sky spectrum along the slit, away from the A and B regions in the slit. In seven cases (PFF-gc091, HHH86-33, HGHH-G204, HH-080, HHH86-39, HGHH-45, and HGHH-G279), we were not able to fully remove the sky emission even when we combined the two techniques, and their final spectra therefore show varying sky contributions.
\par
\begin{figure*}
    \centering
    \includegraphics[width=512pt]{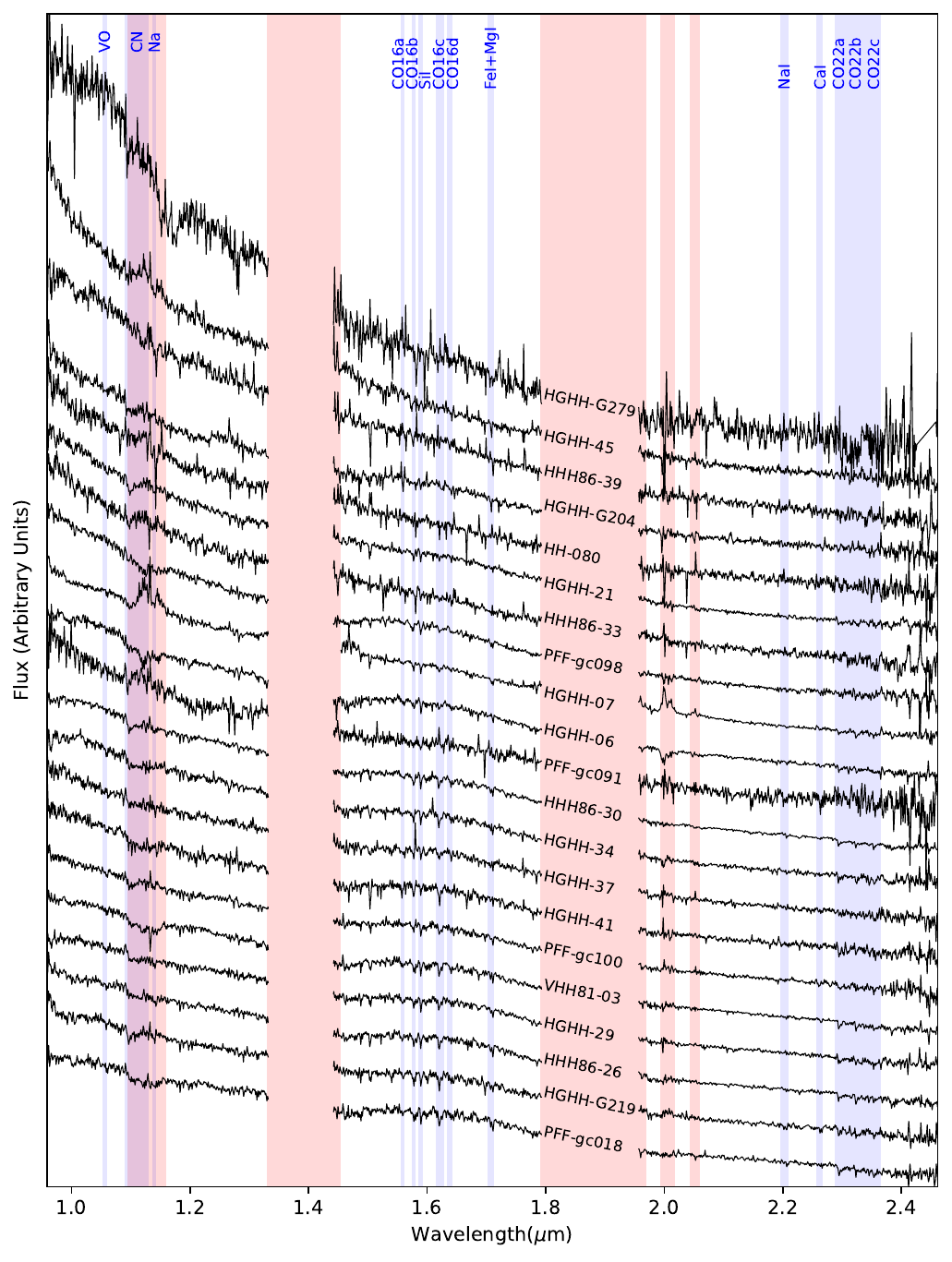}
    \caption{NIR spectra of our sample, normalized at the H band plus a constant, and sorted by J-K colour. Regions with high telluric absorption are masked in red. We also highlight in blue the main stellar absorptions from \citet{Riffel+19} with a sufficient S/N and without an optical counterpart.}
    \label{fig:spec}
\end{figure*}

After the reduction process, we corrected the data for Doppler shift based on the radial velocities shown in Table~\ref{tab:sample}, as well as for Milky Way dust reddening using the \citet{Cardelli+89} law and the \citet{Schlegel+98} extinction maps.  In order to test the consistency of our data, we also derived redshifts for our sample based on H-band data, where most of the stellar absorption lies. These measurements were performed using the code {\sc starlight} \citep{CF+04,CF+05} and E-MILES models \citep{Vazdekis+16}. We found a standard deviation of 30~km~s$^{-1}$ between the optical and NIR measurements, except for two GCs (HGHH-45 and HGHH-G279), for which the method did not converge due to poor S/N and sky residuals. The final rest-frame spectra covering the range 0.96-2.46\,$\mu$m are shown in Fig~\ref{fig:spec}.
%redshift initially taken to be 0.00183

%\begin{sidewaysfigure*}
%    \centering
%    \includegraphics[width=\textwidth]{spec2.pdf}
%    \caption{Continuation of Fig~\ref{fig:spec1} with the second half of our sample.}
%    \label{fig:spec2}
%\end{sidewaysfigure*}

\subsubsection{Photometry}
\label{sec:photometry}

In order to provide a catalogue that is as complete as possible for future studies, we also compiled optical and NIR photometrical points. These values are presented in Table \ref{tab:mags} for the globular clusters in our sample. The ${u}'{g}'{r}'{i}'{z}'$ photometry is from \citet{Taylor+17}. This catalogue comprises $\sim$3200~GCs that were observed with the Dark Energy Camera (DECam) at the Blanco Telescope in Cerro Tololo. 
%Their catalogued photometry and associated uncertainties are listed on Table \ref{tab:mags}.

Situated at Stripe 30, Cen~A was observed in the NIR by the Visible and Infrared Survey Telescope for Astronomy \citep[VISTA][]{Sutherland+15} as part of the VISTA Hemisphere Survey \citep[VHS,][]{McMahon+13}. VHS uses the 4m VISTA telescope, located at ESO Paranal Observatory in Chile, to map the southern hemisphere in at least two bands (J and K$_{s}$) with 5$\sigma$ point-source limiting magnitudes of J$_{\rm{VEGA}}$ = 20.6 and K$_{s,\rm{VEGA}}$ = 18.5 \citep{McMahon+13}.
Using VISTA public data for Cen A, we performed aperture photometry for the 21 GCs in the J and K$_{s}$ bands using an aperture radius of 6 pixels and circular annuli to estimate the local background.
%The local background was estimated and subtracted using circular annuli of inner and outer radii of 12 and 18 pixels, respectively. %Prior to effectively calculating the magnitude, the centroid of the sources is computed and recentering up to 1 pixel in both spatial directions is allowed.
The magnitudes were corrected for foreground Galactic dust extinction using the VHS-DR5 catalogue values, which are based on the Schlegel dust maps \citep{Schlegel+98}. The final magnitudes and associated errors are reported in Table \ref{tab:mags}, where the error is calculated as the square root of the sum in quadrature of the photometric and zero-point errors. The final magnitude values were compared to those from the VHS catalogue to check for consistency. The values agree fairly well, with a median absolute difference of 0.03 mag in the J band and 0.06 in the K$_{s}$ band. To keep the consistency of photometric systems used in Table \ref{tab:mags}, we converted the optical magnitudes into the Vega system using the \textsc{pyphot} Python package \citep{pyphot2022}.
%To keep consistency of photometric systems used in Table \ref{tab:mags} we convert the NIR magnitudes to AB using $J_{\rm{AB}} = J_{\rm{VEGA}} + 0.916$ and $K_{s,\rm{AB}} = K_{s,\rm{VEGA}} + 1.827$\footnote[1]{\url{http://casu.ast.cam.ac.uk/surveys-projects/vista/technical/filter-set}}.

\subsubsection{Data quality}
\label{sec:DQ}
\begin{figure}
    \centering
    \includegraphics[width=\columnwidth]{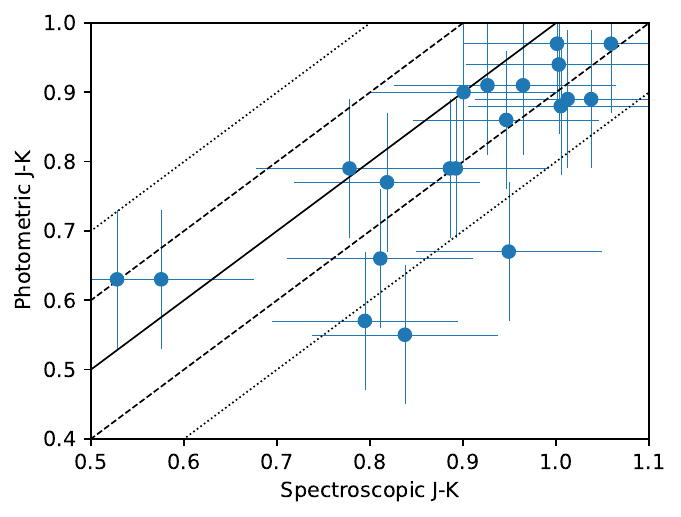}
    \caption{Comparison between J-K colours derived from our spectra with those derived from VISTA photometry. The solid line represents a perfect correlation. The dashed and dotted lines represent a deviation of 0.1 and 0.2 magnitudes from a perfect correlation, respectively. The anomalous/intermediate-age GC HGHH-G279 is not included in this plot because we were unable to derive its K-band photometry from its DECam data.}
    \label{fig:colours}
\end{figure}

Our observations were not performed under a photometric sky, and our flux calibration is therefore not absolute, meaning that the magnitude estimates derived from our spectra should not be directly compared to other works. However, since these observations have been made using cross-dispersed (XD) mode, which obtains the J, H, and K bands at the same time, the colours are much less affected than magnitude zero points because the whole spectra were taken under the same atmosphere \citep[see][for example]{Nikolaev+00}. In order to estimate the uncertainty in the colour measurements, we derived J-H and H-Ks colours in our spectra \citep[calculated based on 2MASS filters,][]{Cohen+03}. These colours are presented in Table~\ref{tab:DQ}. In Fig~\ref{fig:colours}, we compare them with the colours derived from VISTA photometry (Sect~\ref{sec:photometry}). These results indicate that 13 objects are within 0.1 magnitudes of a perfect correlation, 4 objects are between 0.1 and 0.2 magnitudes away from a perfect correlation, and only 3 objects are more than 0.2 magnitudes away. 

We also derived the S/N of the data set over the entire wavelength range. The maximum value achieved in each band is also presented in Table~\ref{tab:DQ}, together with the total integration time on target. The object with the highest S/N of our sample is HHH86-30. It reaches almost 100 at the H band. On the other hand, the object with the lowest S/N is HGHH-G279, which is also the faintest object of our sample. Its S/N peaks at 15 in the J band.
\par

\begin{table}
    \centering
        \caption{Properties of our NIR spectra. Total on-source exposure time, S/N, and NIR colours, as derived from our spectra.}
    \begin{tabular}{lcccccr}
\hline
\hline
ID         & Exp Time & \multicolumn{3}{c}{S/N{\sc max}} & \multicolumn{2}{c}{Colours} \\
           & (mins)   & J & H & K & J-H     &              H-K \\
\hline
HGHH-06    &54   &36  &54  &34   &0.76  & 0.12\\
HGHH-07    &60   &69  &88  &47   &0.76  & 0.12\\
HGHH-21    &72   &50  &61  &33   &0.69  & 0.08\\
HGHH-29    &48   &42  &56  &31   &0.83  & 0.18\\
HGHH-34    &48   &37  &48  &26   &0.77  & 0.14\\
HGHH-37    &42   &30  &40  &22   &0.79  & 0.14\\
HGHH-41    &48   &29  &35  &20   &0.80  & 0.15\\
HGHH-45    &60   &30  &28  &15   &0.46  & 0.06\\
HGHH-G204  &60   &26  &28  &17   &0.69  & 0.11\\
HGHH-G219  &60   &34  &44  &26   &0.81  & 0.22\\
HGHH-G279  &306  &15  &13  &5    &0.30  &-0.05\\
HH-080     &72   &20  &23  &12   &0.70  & 0.12\\
HHH86-26   &36   &50  &66  &39   &0.80  & 0.19\\
HHH86-30   &72   &78  &99  &63   &0.77  & 0.13\\
HHH86-33   &84   &28  &30  &16   &0.64  & 0.14\\
HHH86-39   &60   &24  &26  &13   &0.55  & 0.02\\
PFF-gc018  &48   &39  &53  &30   &0.84  & 0.21\\
PFF-gc091  &90   &16  &21  &10   &0.76  & 0.17\\
PFF-gc098  &60   &46  &58  &27   &0.72  & 0.09\\
PFF-gc100  &66   &43  &61  &28   &0.82  & 0.18\\
VHH81-03   &48   &54  &72  &45   &0.80  & 0.19\\
\hline
    \end{tabular}
    \label{tab:DQ}
\end{table}

%\begin{figure*}
%    \centering
%    \includegraphics[width=500pt]{SNR.pdf}
%    \caption{Signal-to-noise ratio for all objects in our sample.}
%    \label{fig:SNR}
%\end{figure*}

\section{Cen~A LIBERTY properties}
\label{sec:results}

In Fig.~\ref{fig:sample} (panels 2, 3, and 4), we present the correlations between NIR colours derived from our spectra (Sect~\ref{sec:DQ}) and optical H$\beta$ and [MgFe]' indices. These properties show varying correlations, and the strongest is J-H$\times$H-K. 
\par
A special case in our sample is HGHH-G279. Its colours and H$\beta$ are extremely different from the others. The only property that is similar is its [MgFe]', whose value lies within the values of the other objects of our sample. This object was spectroscopically identified as a younger cluster candidate by \citet{Held+02}, and it was photometrically identified by \citet{Peng+04}. B08 reported that this object is well reproduced by SSP models with [Z/H]=-1.2 and 1.7\,Gyr. Its J-H and H-K colours, which are very different from those of the other objects in our sample, are consistent with this scenario. However, due to the low S/N of our spectra, it is currently not possible to perform a spectral index analysis or a full spectral fitting, as these would produce unreliable results. Thus, because this GC is unique, a spectrum with a higher S/N needs to be obtained, which would help us understand the NIR stellar population analysis.
\par
Two other objects also show NIR colours that are different from those of the bulk of the sample: HGHH-45 and HHH86-39, with J-H of 0.46 and 0.55, and H-K of 0.06 and 0.02, respectively. However, their optical indices are similar to those of the other GCs. The fact that these objects show similar optical properties but are different in the NIR, is also of much interest because these differences might be able to help us set better constraints on the models. One possibility to explain these differences is that one of these GCs might have strong hot horizontal branch stars \citep[e.g.][]{Cantiello&Blakeslee07,Chies-Santos+12,Cabrera-Ziri&Conroy22}. Because the NIR is more affected by populations like this, we would be able to spot these differences either by an index analysis or through a full spectral fitting. Future analyses should therefore focus on the cause of these discrepancies. All other objects fall within a similar region that is delimited by J-H$>$0.64 and H-K$>$0.09.

\section{Final remarks}
\label{sec:remarks}
We have presented a NIR (0.96-2.46\,$\mu$m) spectral survey of extragalactic globular clusters, containing 21 GCs from the galaxy Cen~A. These spectra cover H$\beta$ EWs between 0.98 and 4.32 as well as [MgFe]' between 0.23 and 3.76. This data set was observed with SOAR/TripleSpec4 with a spectral resolution of $\sim$3500.
\par
These spectra cover the main NIR stellar absorptions and are suitable for an index analysis and for a full spectral fitting. The S/N of our spectra varies within our sample. It peaks between 15 and 100, depending on the object. The colours are also consistent with the photometrical measurements. In our sample, 65\% of the objects have J-K colours within 0.1 magnitudes of their photometrical counterpart.
\par
Preliminary results show that most of our sample has similar NIR colours as well as optical ages and metallicities. The NIR colours of three objects (HGHH-G279,  HGHH-45, and HHH86-39), on the other hand, differ from those of the rest of our sample. These objects are of great interest to future analyses because their properties are rare among our GCs. Moreover, two of these objects have similar NIR colours but different Lick/IDS indices in the optical, which could indicate a strong difference in their hot horizontal branches.
\par
With the current and upcoming facilities that focus on the infrared, it is very important to develop new techniques that can properly determine ages, metallicities, and general abundances from this wavelength range. This set of spectra provides a unique opportunity to understand the limitations of current models in detail, as well as to develop tools for overcoming these limitations. With this data set in hands, we investigated the zero-point problem in the NIR \citep[Paper~II,][]{Eftekhari+25}, and three future papers are planned to investigate the CO absorptions, the remaining indices, and a full spectral fitting. 
\par
Compared to previous NIR libraries of GCs, such as \citet{Lyubenova+10, Sakari+16}, our library has a similar age and metallicity coverage, but encompasses the whole NIR at once, thus allowing for a wider range of studies. Because our goal is to collect as many GCs in different environments, adding NIR spectra of another galaxy moreover allows for much broader studies of the GC evolution.

\section{Data availability}

The raw data used throughout this paper are already public and can be accessed through the NOIRLab Astro Data Archive (\url{https://astroarchive.noirlab.edu/}). We also intend to make the reduced data fully available for the community in the future. However, since we are still working on two more papers, we can make the data available under reasonable request by e-mail\footnote{Preferably to either the first (\url{luisgdh@gmail.com}) or second author (\url{ana.chies@ufrgs.br}).}.

\begin{acknowledgements}
We thank the anonymous referee for their careful reading of our manuscript and their many
insightful comments and suggestions. We acknowledge Chris Usher for useful discussions. LGDH acknowledges support by National Key R\&D Program of China No.2022YFF0503402, and National Natural Science Foundation of China (NSFC) project number E345251001.  ACS acknowledges funding from the Conselho Nacional de Desenvolvimento Científico e Tecnológico (CNPq), the Rio Grande do Sul Research Foundation (FAPERGS) and the Chinese Academy of Sciences (CAS) President's International Fellowship Initiative (PIFI) through grants CNPq-11153/2018-6, CNPq-314301/2021-6, FAPERGS/CAPES 19/2551-0000696-9, E085201009. RR acknowledges support from the Fundaci\'on Jes\'us Serra and the Instituto de Astrof{\'{i}}sica de Canarias under the Visiting Researcher Programme 2023-2025 agreed between both institutions. RR, also acknowledges support from the ACIISI, Consejer{\'{i}}a de Econom{\'{i}}a, Conocimiento y Empleo del Gobierno de Canarias and the European Regional Development Fund (ERDF) under grant with reference ProID2021010079, and the support through the RAVET project by the grant PID2019-107427GB-C32 from the Spanish Ministry of Science, Innovation and Universities MCIU. This work has also been supported through the IAC project TRACES, which is partially supported through the state budget and the regional budget of the Consejer{\'{i}}a de Econom{\'{i}}a, Industria, Comercio y Conocimiento of the Canary Islands Autonomous Community. RR also thanks to Conselho Nacional de Desenvolvimento Cient\'{i}fico e Tecnol\'ogico  ( CNPq, Proj. 311223/2020-6,  304927/2017-1 and  400352/2016-8), Funda\c{c}\~ao de amparo \`{a} pesquisa do Rio Grande do Sul (FAPERGS, Proj. 16/2551-0000251-7 and 19/1750-2), Coordena\c{c}\~ao de Aperfei\c{c}oamento de Pessoal de N\'{i}vel Superior (CAPES, Proj. 0001). AEL acknowledges the support from Coordena\c{c}\~ao de Aperfei\c{c}oamento de Pessoal de N\'{i}vel Superior (CAPES) in the scope of the Program CAPES-PrInt, process number 88887.837405/2023-00 and CAPES-PROEX fellowship, process number 88887.513351/2020-00. EE and AV acknowledge support from grant PID2021-123313NA-I00 and
PID2022-140869NB-I00 from the Spanish Ministry of Science, Innovation and Universities MCIU, as well as the grant POKEBOWL PID2021-123313NA-I00 from the MCIN/AEI and the European Regional Development Fund (ERDF).
\end{acknowledgements}

%-------------------------------------------------------------------

\bibliographystyle{aa}
\bibliography{luisgdh} 

\begin{appendix}

\section{Sample properties and photometry}

In Tab~\ref{tab:sample}, we present the sample used throughout this paper, as well as their main properties. In Tab~\ref{tab:mags}, we show optical and NIR photometrical points, derived from DECam/Blanco and the VISTA survey, respectively.

\begin{table*}
    \centering
            \caption{Sample properties}
    \begin{tabular}{lccccHHHHHcr}
\hline
\hline
ID          & R.A (J2000) & Dec. (J2000) & V (Mag) & $R_{V}$ (km/s) &  [M/H]   & $\sigma_{[M/H]}$ & Age(Gy) & $\sigma_{Age}$ & SN &  H$\beta$ (\r{A}) & [MgFe]' (\r{A})\\
\hline
HGHH-06     & 13 25 22.19 & -43 02 45.6  & 17.2 & 790$\pm$44   &  -1.100  & 0.140            & 10.000  & 1.522          & 43 & 2.61 & 1.16\\
HGHH-07     & 13 26 05.41 & -42 56 32.4  & 17.2 & 593$\pm$51   &  -1.025  & 0.121            & 11.885  & 2.581          & 92 & 1.92 & 1.89\\
HGHH-21     & 13 25 52.74 & -43 05 46.4  & 17.9 & 495$\pm$38   &  -1.100  & 0.138            & 11.220  & 2.456          & 50 & 1.90 & 1.74\\
HGHH-29     & 13 24 40.39 & -43 18 05.3  & 18.1 & 743$\pm$34   &  -0.450  & 0.121            & 11.220  & 2.454          & 65 & 1.75 & 2.89\\
HGHH-34     & 13 25 40.61 & -43 21 13.6  & 18.1 & 676$\pm$27   &  -0.450  & 0.114            &  8.414  & 2.430          & 57 & 2.00 & 2.69\\
HGHH-37     & 13 26 10.58 & -42 53 42.7  & 18.4 & 620$\pm$28   &  -0.700  & 0.137            & 11.885  & 2.444          & 71 & 1.77 & 2.29\\
HGHH-41     & 13 24 38.98 & -43 20 06.4  & 18.6 & 394$\pm$27   &  -0.325  & 0.141            &  6.310  & 2.491          & 59 & 1.66 & 2.78\\
HGHH-45     & 13 25 34.25 & -42 56 59.1  & 19.0 & 618$\pm$56   &  -0.850  & 0.255            &  3.548  & 3.331          & 39 & 2.25 & 1.67\\
HGHH-G204   & 13 25 46.99 & -43 02 05.4  & 18.3 & 705$\pm$36   &  -2.000  & 0.252            & 11.885  & 0.962          & 35 & 2.61 & 0.23\\
HGHH-G219   & 13 25 17.31 & -42 58 46.6  & 18.8 & 534$\pm$28   &  -0.025  & 0.138            &  3.981  & 1.525          & 51 & 1.22 & 3.76\\
HGHH-G279   & 13 24 56.27 & -43 03 23.4  & 19.5 & 338$\pm$54   &  -1.175  & 0.415            &  1.679  & 1.125          & 34 & 4.32 & 0.47\\
HH-080      & 13 23 38.33 & -42 46 22.8  & 17.8 & 497$\pm$33   &  -1.575  & 0.070            & 10.000  & 0.534          & 71 & 2.35 & 0.83\\
HHH86-26    & 13 26 15.27 & -42 48 29.4  & 18.1 & 405$\pm$24   &  -0.275  & 0.083            & 11.885  & 1.166          & 41 & 1.34 & 3.12\\
HHH86-30    & 13 24 54.35 & -42 53 24.8  & 17.3 & 811$\pm$32   &  -0.375  & 0.130            &  5.012  & 1.830          & 69 & 2.00 & 2.39\\
HHH86-33    & 13 25 16.26 & -42 50 53.3  & 18.5 & 522$\pm$39   &  -1.350  & 0.113            & 10.593  & 1.939          & 69 & 2.15 & 1.24\\
HHH86-39    & 13 26 42.03 & -43 07 44.8  & 17.4 & 249$\pm$59   &  -0.475  & 0.052            & 11.885  & 0.000          & 79 & 0.98 & 2.92\\
PFF-gc018   & 13 24 47.10 & -43 06 01.7  & 18.9 & 531$\pm$29   &  -0.050  & 0.095            &  9.441  & 2.245          & 66 & 1.54 & 3.52\\
PFF-gc091   & 13 26 21.14 & -43 42 24.6  & 19.3 & 669$\pm$57   &  -1.075  & 0.186            &  5.309  & 2.895          & 28 & --   & --  \\
PFF-gc098   & 13 26 53.94 & -43 19 17.7  & 18.3 & 650$\pm$37   &  -0.725  & 0.242            &  4.732  & 3.593          & 41 & 2.11 & 1.78\\
PFF-gc100   & 13 27 03.41 & -42 27 17.2  & 18.4 & 527$\pm$31   &  -0.475  & 0.167            &  5.957  & 3.243          & 64 & 2.11 & 2.53\\
VHH81-03    & 13 24 58.21 & -42 56 10.0  & 17.7 & 591$\pm$37   &  -0.200  & 0.153            &  5.309  & 2.161          & 48 & 1.83 & 2.80\\
    \end{tabular}
    \begin{tablenotes}
    \item Notes: Main properties of our sample: (1) Name; (2 and 3) equatorial coordinates; (4) V-band magnitude; (5) radial velocity with uncertainty (Km/s) derived by B08; (6 and 7) H$\beta$ and [MgFe]' indices derived by B08.
    \end{tablenotes}
    \label{tab:sample}
\end{table*}

\begin{table*}
    \centering
        \caption{Photometry and associated uncertainties for the GC sample in the Vega photometric system.}
    \begin{tabular}{cccccccc}
\hline
\hline
ID        & ${u}'$                & ${g}'$                & ${r}'$                & ${i}'$                & ${z}'$                & J                & K$_{s}$               \\ \hline
HGHH-06   & $-$              & $-$              & $-$              & $-$              & $-$              & 15.04 $\pm$ 0.11 & 14.25 $\pm$ 0.12 \\
HGHH-07 & 18.66 $\pm$ 0.08 & 17.49  $\pm$ 0.03 & 16.64  $\pm$ 0.08 & 15.95  $\pm$ 0.01 & 15.88 $\pm$ 0.01 & 15.21 $\pm$ 0.04 & 14.42 $\pm$ 0.07 \\
HGHH-21   & 19.92 $\pm$ 0.08 & 18.60 $\pm$ 0.03  & 17.60 $\pm$ 0.08 & 16.87 $\pm$ 0.01 & 16.69 $\pm$ 0.02 & 15.85 $\pm$ 0.06 & 15.06 $\pm$ 0.10 \\
HGHH-29   & 20.56 $\pm$ 0.08 & 18.90 $\pm$ 0.03 & 17.81 $\pm$ 0.08 & 17.00 $\pm$ 0.01 & 16.73 $\pm$ 0.02 & 15.78 $\pm$ 0.05 & 14.89 $\pm$ 0.11 \\
HGHH-34   & 20.76 $\pm$ 0.08 & 19.11 $\pm$ 0.03 & 18.15 $\pm$ 0.08 & 17.28 $\pm$ 0.01 & 17.02 $\pm$ 0.02 & 16.09 $\pm$ 0.08 & 15.18 $\pm$ 0.11 \\
HGHH-37   & 20.65 $\pm$ 0.08 & 19.17 $\pm$ 0.03 & 18.10 $\pm$ 0.08 & 17.39 $\pm$ 0.01 & 17.13 $\pm$ 0.02 & 16.29 $\pm$ 0.08 & 15.43 $\pm$ 0.11 \\
HGHH-41   & 20.41 $\pm$ 0.08 & 18.92 $\pm$ 0.03 & 17.82 $\pm$ 0.08 & 17.22 $\pm$ 0.02 & 16.99 $\pm$ 0.02 & 16.22 $\pm$ 0.08 & 15.31 $\pm$ 0.14 \\
HGHH-45   & $-$              & $-$              & $-$              & $-$              & $-$              & 17.13 $\pm$ 0.22 & 16.5 $\pm$ 0.5   \\
HGHH-G204 & 19.57 $\pm$ 0.08 & 18.59 $\pm$ 0.03 & 17.83 $\pm$ 0.08 & 17.21 $\pm$ 0.02 & 17.07 $\pm$ 0.02 & 16.49 $\pm$ 0.13 & 15.83 $\pm$ 0.23 \\
HGHH-G219 & 20.89 $\pm$ 0.08 & 19.16 $\pm$ 0.03 & 18.18 $\pm$ 0.08 & 17.39 $\pm$ 0.02 & 17.14 $\pm$ 0.02 & 16.39 $\pm$ 0.17 & 15.50 $\pm$ 0.24 \\
HGHH-G279 & 20.09 $\pm$ 0.08 & 19.52 $\pm$ 0.04 & 19.10 $\pm$ 0.08 & 18.71 $\pm$ 0.02 & 18.72 $\pm$ 0.03 & 18.5 $\pm$ 0.7   & $-$              \\
HH-080    & 20.22 $\pm$ 0.08 & 19.12 $\pm$ 0.03 & 18.18 $\pm$ 0.08 & 17.52 $\pm$ 0.02 & 17.34 $\pm$ 0.02 & 16.65 $\pm$ 0.15 & 16.1 $\pm$ 0.4   \\
HHH86-26  & 20.34 $\pm$ 0.08 & 18.61 $\pm$ 0.03 & 17.54 $\pm$ 0.08 & 16.92 $\pm$ 0.01 & 16.47 $\pm$ 0.01 & 15.58 $\pm$ 0.05 & 14.61 $\pm$ 0.10 \\
HHH86-30  & 19.03 $\pm$ 0.08 & 17.60 $\pm$ 0.03 & 16.65 $\pm$ 0.08 & 15.90 $\pm$ 0.01 & 15.76 $\pm$ 0.01 & 15.02 $\pm$ 0.03 & 14.12 $\pm$ 0.05 \\
HHH86-33  & 19.78 $\pm$ 0.08 & 18.80 $\pm$ 0.03 & 17.98 $\pm$ 0.08 & 17.39 $\pm$ 0.02 & 17.28 $\pm$ 0.02 & 16.79 $\pm$ 0.12 & 16.22 $\pm$ 0.32 \\
HHH86-39  & 19.05 $\pm$ 0.08 & 17.78 $\pm$ 0.03 & 16.87 $\pm$ 0.08 & 16.32 $\pm$ 0.01 & 16.18 $\pm$ 0.01 & 15.73 $\pm$ 0.04 & 15.10 $\pm$ 0.13 \\
PFF-gc018 & 21.14 $\pm$ 0.08 & 19.33 $\pm$ 0.03 & 18.25 $\pm$ 0.08 & 17.48 $\pm$ 0.02 & 17.16 $\pm$ 0.02 & 16.28 $\pm$ 0.09 & 15.31 $\pm$ 0.14 \\
PFF-gc091 & 21.18 $\pm$ 0.08 & 20.01 $\pm$ 0.04 & 19.00 $\pm$ 0.08 & 18.34 $\pm$ 0.02 & 18.11 $\pm$ 0.02 & 17.47 $\pm$ 0.29 & 16.8 $\pm$ 0.5   \\
PFF-gc098 & 20.35 $\pm$ 0.08 & 19.01 $\pm$ 0.03 & 17.95 $\pm$ 0.08 & 17.27 $\pm$ 0.01 & 17.07 $\pm$ 0.02 & 16.29 $\pm$ 0.08 & 15.52 $\pm$ 0.15 \\
PFF-gc100 & 20.68$\pm$ 0.08  & 19.14 $\pm$ 0.03 & 18.07 $\pm$ 0.08 & 17.30 $\pm$ 0.01 & 17.02 $\pm$ 0.02 & 16.15 $\pm$ 0.08 & 15.27 $\pm$ 0.17 \\
VHH81-03  & 19.65 $\pm$ 0.08 & 18.09 $\pm$ 0.03 & 17.14 $\pm$ 0.08 & 16.36 $\pm$ 0.01 & 16.19 $\pm$ 0.01 & 15.34 $\pm$ 0.03 & 14.40 $\pm$ 0.05 \\ \hline
    \end{tabular}
    \begin{tablenotes}
    \item Notes:  The photometry for the DECam filters ${u}'{g}'{r}'{i}'{z}'$ was taken from the catalogue of \citet{Taylor+17}, and the NIR magnitudes from VISTA were obtained via aperture photometry (see text for details of the method).
    \end{tablenotes}
    \label{tab:mags}
\end{table*}

\end{appendix}

\end{document}